\documentclass{aastex}          

\newcommand{\diff}[2]{\frac{\partial #1}{\partial #2}}
\newcommand{\gesim}{\:\raisebox{-0.9ex}{$\stackrel{\displaystyle >}{\sim}$}\:}
\newcommand{\mapright}[1]{\smash{\mathop{\hbox to 1cm{\rightarrowfill}}\limits^{#1}}}

\begin{document}

\title{Monte Carlo Study on Distortion of the Space-Dimension in COBE Monopole Data}
\shorttitle{Distortion of Space-Dimension in COBE Data}
\shortauthors{Biyajima \& Mizoguchi}

\author{Minoru Biyajima}
\affil{Department of Physics, Shinshu University, Matsumoto 390-8621, Japan}
\email{biyajima@azusa.shinshu-u.ac.jp}
\and
\author{Takuya Mizoguchi}
\affil{Toba National College of Maritime Technology, Toba 517-8501, Japan}
\email{mizoguti@toba-cmt.ac.jp}

\begin{abstract}
A concise explanation of studies on distortion of space-time dimension is briefly introduced. Second we obtain the limits (i.e., bounded values) of the dimensionless chemical potential $\mu$, the Sunyaev--Zeldovich (SZ) effect $y$ and distortion of the space-dimension $\varepsilon$ by Monte Carlo (MC) analysis of the parameter set ($T$, $d=3+\varepsilon$, $\mu$, and $y$) in cosmic microwave data assuming that the SZ effect is positive $(y>0)$. In this analysis, the magnitude of the space-dimension $d$ with distortion of the  space-dimension $\varepsilon$ is defined by $d=3+\varepsilon$. The limits of $\mu$ and $y$ are determined as $|\mu| < 9\times 10^{-5}$ ($2\sigma$) ($\mu = (-3.9\pm 2.6)\times 10^{-5}$ ($\sigma$)), $|y| < 5\times 10^{-6}$ ($2\sigma$) ($y = (2.0\pm 1.4)\times 10^{-6}$ ($\sigma$)), while the distortion of the space-dimension is $|\varepsilon| < 6\times 10^{-5}$ ($2\sigma$) ($\varepsilon = (-0.78\pm 2.50)\times 10^{-5}$ ($\sigma$)). The magnitudes of these three estimated limits are ordered as $|\mu|\gesim |\varepsilon| > |y|$. The estimated limit of $|y| < 5\times 10^{-6}$ appears to be related to re-ionization processes occurring at redshift $z_{ri}\sim 10$. We also present data analysis assuming a relativistic SZ effect.
\end{abstract}

\keywords{cosmic background radiation; cosmological parameters; space-dimension}

\section{\label{sec1}Introduction}
First, we briefly introduce the essential background to topic of this paper. About one century ago, \citet{Ehrenfest1917} published an interesting paper titled ``In What Way Does It Become Manifest in the Fundamental Law of Physics that Space Has Three Dimensions?''. This paper remains actively discussed even today (see a paper ``dimensionality'' by \citet{Barrow1983}). About 40 years ago, in quantum field theory, a method of dimensional regularization, permitting a distortion of space-time dimension $D-4=\varepsilon$, was proposed by \citet{tHooft1972}, \citet{Bollini1972} and \citet{Ashmore1972}. This method is equivalent to assigning an infinitesimal fictitious mass $\lambda_{\rm min}$ to a photon or introducing a large momentum cut-off $\Lambda$ in the momentum integration of the Feynman diagram ($2/\varepsilon$ corresponds to $\log (m_e/\lambda_{\rm min})^2$ or $\log (\Lambda/m_e)^2$, where $m_e$ is the mass of an electron). These methods are important in calculations of physical quantities such as the self-energy and the anomalous magnetic moment of electron (and muon), suggesting that non-integer values of $D=3 + \varepsilon + 1$ may have a feasible physical interpretation.

Shortly after 't~Hooft and Veltman, and their contemporaries proposed their method in 1972, \citet{Mandelbrot1977} conceptualized the fractal dimension of space-time. Inspired by his book, several researchers have investigated fractal dimensions in the natural sciences \citep{Zeilinger1985,Jarlskog1985,Muller1986,Grassi1986,Torres1988}. A range of estimated physical quantities is summarized in Table \ref{tab-1}. Among these, the phase space of fractal $d_H$-dimension (named the Hausdorff dimension) is given by \citet{Mandelbrot1977} based on \citet{Hausdorff1918} 
\begin{eqnarray}
\gamma(d_H) = \frac{[\Gamma (1/2)]^{d_H}}{\Gamma (1+d_H/2)},
\label{eq-1}
\end{eqnarray}
where $\Gamma$ is the gamma function, and Eq. (\ref{eq-1}) can be multiplied by a factor $(\nu/c)^d$ in actual calculations. (See Eq. (\ref{eq2}) below.)

\begin{table*}[htbp]
  \caption{\label{tab-1}Various estimates of distortion parameters in the space-time dimension ($D = d + 1$ or $d = 3 + \varepsilon$). }
  \vspace{-2mm}
  \begin{center}
  \begin{tabular}{lll}
  \hline
  Authors (Refs.) & Measured (or estimation) quantities\hspace*{-2mm} & $|D-4|$ or $|d-3|=|\varepsilon|$\\
  \hline
  \citet{Zeilinger1985} 
  & Anomalous ($g-2$) electron & $|d_H -4|$\\
  & factor / QED with Hausdorff & $= (5.3\pm 2.5)\times 10^{-7}$\\
  & dimension $d_H$\\
  \hline
  \citet{Jarlskog1985} \hspace*{-2mm}
  & Perihelion of planet Mercury & $|\varepsilon|_M < 1.7\times 10^{-9}$\\
  & Binary PSR 1913+16 & $|\varepsilon|_{PSR} < 1.7\times 10^{-9}$\\
  \hline
  \citet{Muller1986} 
  & Lamb shift of hydrogen & $|\varepsilon| \approx 10^{-11}$\\
  \hline\hline
  \citet{Grassi1986} 
  & Eq.~(\ref{eq2}) & $|\varepsilon| < 2\times 10^{-2}$\\
  \hline
  \citet{Torres1988} 
  & Eq.~(\ref{eq2}) and Stefan-Boltzmann & $|\varepsilon| < 10^{-3}$\\
  & constant including $\varepsilon$\\
  \hline
  \end{tabular}
  \end{center}
  \vspace{-3mm}
\end{table*}

Here, we focus on a paper by \citet{Grassi1986}. The Planck distribution, i.e., the black-body radiation law in $d = 3 + \varepsilon$ dimensions (where $d$ and $\varepsilon$ denote the space-dimension and its distortion based on Eq. (\ref{eq-1}), respectively) is given by
\begin{eqnarray}
U(T,\: \nu,\: \varepsilon) = N_c(d)\frac{(\nu/c)^d}{e^x-1},
\label{eq2}
\end{eqnarray}
where $x=h\nu/k_BT$, $N_c(d) = h\gamma (d)\cdot d(d-1) = 2hc\,\pi^{d/2}(d-1)/[c\Gamma(d/2)]$. The quantities $h$, $\nu$, $k_B$, $c$, and $T$ are Planck's constant, the light frequency, Boltzmann's constant, speed of light, and temperature, respectively. \citet{Grassi1986} suggested that upper limit may be placed on the distortion of space-dimension, such that $|d - 3| < 0.02$ (Note that their investigation was published eight years earlier than the NASA COBE data \citep{Mather1994}). 

Measurements by NASA's COBE satellite revealed that our universe is almost homogenously filled with cosmic microwave radiation at 2.725 K. \citet{Caruso2009} analyzed the \citet{COBE2005} data (where FIRAS denotes Far Infrared Absolute Spectrophotometer) in terms of the following formula:
\begin{eqnarray}
U(T,\: \nu,\: \varepsilon) = N_c\frac{4\pi}c\frac{(\nu/c)^d}{e^x-1},
\label{eq-3}
\end{eqnarray}
where $N_c$ is assumed constant. The results of their analysis are summarized in Table \ref{tab1}. Our recent analyses of the same data using Eq.~(\ref{eq2}) \citep{Biyajima2012} are also depicted in Table \ref{tab1}. As evident from the table, the analyses yield significantly different results, including the signs of $\varepsilon$. These differences are contradicting in spite fact that both studies used the same data, a similar formulation and the CERN MINUIT program. After recalculating our results (listed in the third row of Table \ref{tab1}), we are able to attribute these discrepancies to differences in the numerical values of $hc/k_B$ \citep{Mohr2012,Biyajima2012}.

Second, we note that the COBE data contain two effects, namely, the chemical potential $\mu$, and the Sunyaev--Zeldovich (SZ) effect $y$ \citep{Zeldovich1969,Sunyaev1970} in the black-body radiation spectra. As a next step, it is worthwhile to analyze the COBE data in terms of these effects, in addition to the distortion of the space-dimension $\varepsilon$. 

As is well known, the COBE monopole data contain the following residual spectrum, given by
\begin{eqnarray}
&\!\!\!\!\!\!&\!\!\!\!\!\! [{\rm COBE\ residual\ spectrum}]\nonumber\\
&\!\!\!\!\!\!&\!\!\!\!\!\! \equiv\ [{\rm monopole\ data}]\ - [U_{\rm Planck}\ (T=2.725\ {\rm K})] 
\label{eq3}
\end{eqnarray}
The monopole spectrum is analyzed by means of the Bose--Einstein distribution, which involves the dimensionless chemical potential $\mu$, \citep{Fixsen2002}
\begin{eqnarray}
U_{\rm BE}(T,\: \nu,\: \mu) = C_B\frac{(\nu/c)^3}{e^{(x+\mu)}-1},
\label{eq4}
\end{eqnarray}
where $C_B = 8\pi h$. Moreover, it is necessary to include for the effect of inverse Compton scattering $e^- + \gamma \to e^- + \gamma$, described by 
\begin{eqnarray}
U_{\rm SZ}(T,\: x,\: y) = C_B\frac{(\nu/c)^3yxe^x}{(e^x-1)^2}\left(x\coth \frac x2-4\right).
\label{eq5}
\end{eqnarray}
Here, the SZ effect $y$ \citep{Zeldovich1969,Sunyaev1970} is the parameter for inverse Compton scattering, defined by
\begin{eqnarray}
y \equiv \int dl n_e \sigma_T \frac{k_BT_e}{m_ec^2}
\label{eq6},
\end{eqnarray}
where $l$, $n_e$, $\sigma_T$, and $T_e$ denote the size of the high-temperature region in the Universe, number density of electrons, the cross section of Thomson scattering, and the temperature of electrons, respectively. In this paper, we focus on the positivity of the SZ effect $y$.

To account for the distortion of the spatial dimension in the COBE monopole data, we adopt the following formula, in which $C_B\,(\nu/c)^3$ in the above equation is replaced by $N_c(d)\,(\nu/c)^d$,
\begin{eqnarray}
&\!\!\!\!\!\!&\!\!\!\!\!\!U(T,\: \nu,\: \varepsilon,\: \mu,\: y) = N_c(d)\,\left(\frac{\nu}c\right)^d\nonumber\\
&&\times\left[\frac{1}{e^{(x+\mu)}-1} + \frac{yxe^x}{(e^x-1)^2}\left(x\coth \frac x2-4\right)\right] 
\label{eq7}
\end{eqnarray}
Equation (\ref{eq7}) forms the basis of our investigation on the distortion of the space-dimension in the COBE monopole data.

\begin{table*}
\begin{center}
\caption{\label{tab1}Analysis of COBE monopole data using Eqs. (\ref{eq2}) and (\ref{eq-3})}
\begin{tabular}{@{}cccccc@{}}
\tableline\tableline
& $N_c$ or $2hc$ & $T$ & $\varepsilon$ & $\chi^2$ & inputs for $hc/k_B$\\
& (MJy cm$^3$ sr$^{-1}$) & (K) & ($\times 10^{-6}$) & & (cm K$^{-1}$)\\
\tableline
Eq.~(\ref{eq2}) & $39.7289$ & $2.7250\pm 0.0001$ & $2.9\pm 24.6$  & $45.1$ & $1.43878$\tablenotemark{a}\\
Eq.~(\ref{eq-3}) & $39.73\pm 0.002$ & $2.726\pm 3\times 10^{-5}$ & $-9.6\pm 0.1$ & $45.0$ & 1.43939\tablenotemark{b}\\
\tableline\tableline
\multicolumn{6}{c}{Our recalculation of Eq.~(\ref{eq-3}) using $hc/k_B =1.43878$ and $N_c =$ const.}\\
Eq.~(\ref{eq-3}) & $39.731\pm 0.003$ & $2.7251\pm 0.0001$ & $-53\pm 106$ & $44.8$ & 1.43878\tablenotemark{a}\\
\tableline
\end{tabular}
\tablenotetext{\rm a}{$h = 6.6260696\times 10^{-27}$ [erg s], $c = 2.9979246\times 10^{10}$ [cm s$^{-1}$], and $k_B = 1.3806488\times 10^{-16}$ [erg K$^{-1}$]~\citep{Mohr2012} are used \citep{Biyajima2012}.}
\tablenotetext{\rm b}{Communication with Caruso and Oguri.}
\tablecomments{The difference between the results from Eqs. (\ref{eq2}) and (\ref{eq-3}) is attributed to different numerical values of $hc/k_B$. In the third row, Eq. (\ref{eq-3}) is recalculated using $hc/k_B = 1.43878$ (cm K$^{-1}$), assuming $N_c = constant$.}
\end{center}
\end{table*}

This paper is divided into several sections. In Section \ref{sec2}, we apply Eq. (\ref{eq7}), including and excluding $\varepsilon$, to the COBE monopole data and estimate the physical quantities ($T$, $\varepsilon$, $\mu$, $y$) and $\chi^2$. Assuming these estimates, the COBE data are analyzed at fixed temperatures by the Monte Carlo (MC) method. Average parameter values are estimated for allowed combinations of parameter ensembles. Section \ref{sec3} is denoted to the COBE residual spectrum for these combinations of parameter ensembles. Concluding remarks and discussion, including our analyses of relativistic formula of Eq. (\ref{eq5}) in \citet{Itoh1998,Challinor1998} and an analysis of the dipole spectrum derived from Eq. (\ref{eq2}), are presented in Section \ref{sec4}.

\section{\label{sec2}Analyses of COBE Monopole Data in Terms of the Chemical Potential $\mu$ and the SZ Effect $y$}
\subsection{\label{sec2-1}Analysis of Monopole Data by Equation (\ref{eq7})}
\begin{figure}[h]
\begin{center}
\includegraphics[height=55mm]{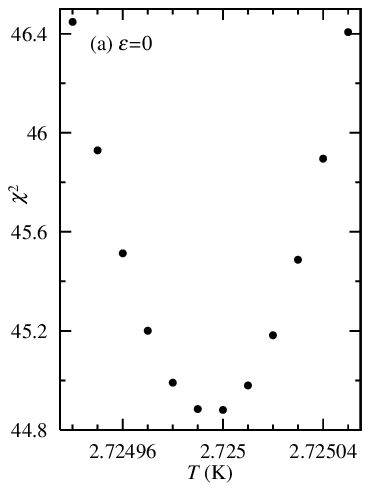}
\includegraphics[height=55mm]{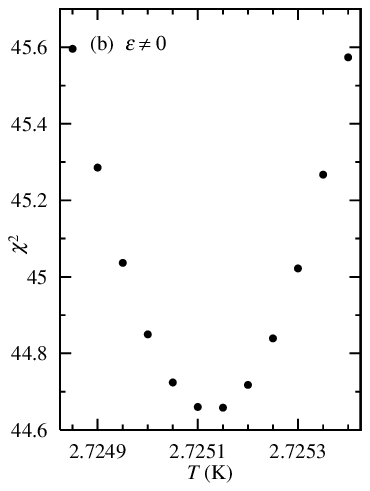}
\caption{\label{fig1}$\chi^2$ as a function of fixed temperature $T$. (a) $\varepsilon = 0$ and (b) $\varepsilon \ne 0$}
\end{center}
\end{figure}

Applying Eq. (\ref{eq7}) to the COBE monopole data \citep{Fixsen2002,COBE2005}, we obtain the results listed in the upper sections of Table \ref{tab2}. From these results, we calculate two parameter sets ($\varepsilon = 0$, $\mu$, $y$) and ($\varepsilon \ne 0$, $\mu$, $y$) at fixed temperatures. The $\chi^2$ widths ($\Delta\chi^2$) of the allowed parameter sets are then estimated from the center-column parameters (listed in the central columns of Table \ref{tab2}) and Fig. \ref{fig1}. From the widths $\Delta\chi^2$ and the standard deviations in the parameters ($\delta\varepsilon$, $\delta\mu$, $\delta y$), we can estimate the distributions of parameters ($\varepsilon$, $\mu$, $y$) at fixed temperatures using the MC method. More details are provided in the following sub-section.
\begin{table*}
\begin{center}
\caption{\label{tab2}Analysis of COBE monopole data by means of Equation (\ref{eq7})}
\begin{tabular}{@{}crrrr@{}}
\tableline\tableline
$T$ (K) & $\varepsilon\;(\times 10^{-5})$ & $\mu\;(\times 10^{-5})$ & $y\;(\times 10^{-6})$ & $\chi^2$\\
\tableline
$2.72501\pm 0.00002$ & --- & $-1.1\pm 3.2$ & --- & $45.0$\\
$2.72504\pm 0.00008$ & $-3.1\pm 6.2$ & $-4.9\pm 8.3$ & --- & $44.7$\\
$2.72500\pm 0.00004$ & --- & $-2.6\pm 5.6$ & $1.6\pm 4.8$ & $44.9$\\
$2.72513\pm 0.00023$ & $-7.6\pm 13.4$ & $-6.9\pm 9.5$ & $-3.7\pm 10.4$ & $44.7$\\
\tableline\tableline
$T=$ fixed\\
$2.72495$ & --- & $-8.3\pm 1.0$ & $5.8\pm 2.6$ & $45.93$\\
$2.72498$ & --- & $-4.6\pm 1.0$ & $3.0\pm 2.6$ & $44.99$\\
$2.72500$ & --- & $-2.1\pm 1.0$ & $1.2\pm 2.6$ & $44.88$\\
$2.72502$ & --- & $0.4\pm 1.0$ & $-0.6\pm 2.6$ & $45.18$\\
$2.72504$ & --- & $2.9\pm 1.0$ & $-2.5\pm 2.6$ & $45.90$\\
\tableline
$T=$ fixed\\
$2.72490$ & $5.2\pm 2.5$   & $-0.4\pm 6.9$  & $5.8\pm 3.4$   & $45.29$\\
$2.72500$ & $-0.4\pm 2.5$  & $-3.2\pm 6.9$  & $1.6\pm 3.4$   & $44.85$\\
$2.72510$ & $-6.1\pm 2.5$  & $-6.2\pm 7.0$  & $-2.6\pm 3.4$  & $44.66$\\
$2.72520$ & $-11.8\pm 2.5$ & $-9.0\pm 6.9$  & $-6.8\pm 3.4$  & $44.72$\\
$2.72535$ & $-20.3\pm 2.5$ & $-13.4\pm 6.9$ & $-13.2\pm 3.4$ & $45.27$\\
\tableline
\end{tabular}
\end{center}
\end{table*}

\subsection{\label{sec2-2}Application of MC Method to the Parameter-Set ($T$, $\mu$, $y$) (Case $\varepsilon = 0$)}
\begin{figure}[h]
\begin{center}
\includegraphics[height=58mm]{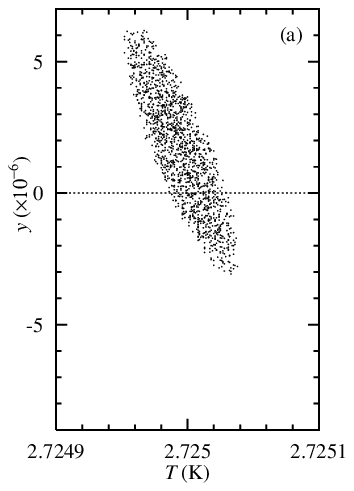}
\includegraphics[height=58mm]{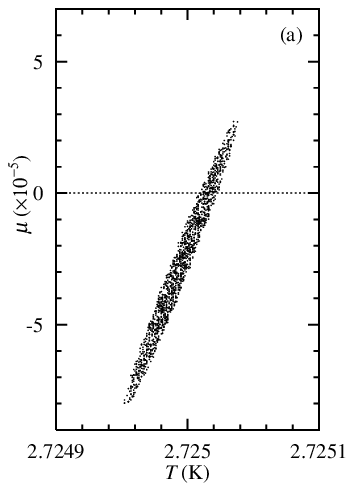}\\
\includegraphics[height=58mm]{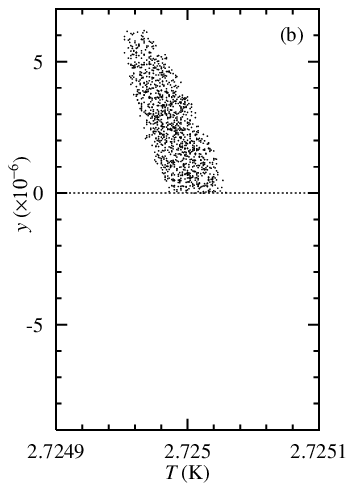}
\includegraphics[height=58mm]{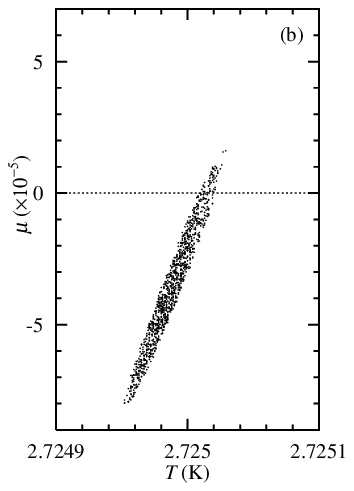}
\end{center}
\caption{\label{fig2}Parameter ensemble ($T,\; \mu,\ y$) with $\chi^2<\chi_{\rm min}^2 (44.87) + \Delta \chi^2 (1.00)$. Number of events generated is $3\times 10^4$ (30k). (a) $N = 1529$. (b) $N_{y>0} = 1142$.}
\end{figure}

The random variables are introduced as follows: 
\begin{eqnarray}
T: &\!\!\!\!\!&\!\!\!\!\! T \pm \delta T \to T + \delta T\times U(-1,\ 1)
\nonumber\\
\mu: &\!\!\!\!\!&\!\!\!\!\! \mu \pm \delta \mu \to \mu + \delta \mu\times U(-1,\ 1)
\nonumber\\
y: &\!\!\!\!\!&\!\!\!\!\! y \pm \delta y \to y + \delta y\times U(-1,\ 1)
\label{eq8}\\
\chi^2: &\!\!\!\!\!&\!\!\!\!\! \chi^2 < \chi_{\rm min}^2 + \Delta \chi^2
\nonumber\\
\Delta \chi^2: &\!\!\!\!\!&\!\!\!\!\! {\rm depending\ on\ the\ temperature\ interval}\ T \pm \delta T.
\nonumber 
\end{eqnarray}
where $U(-1,\ 1)$ are uniform random numbers in $(-1,\ 1)$ (c.f. \citet{Mizoguchi2001}). The resulting $\chi^2$ values are presented in Fig. \ref{fig1}.  The parameter distributions obtained from our analysis are shown in Fig. \ref{fig2}.  Assuming that the SZ effect $y$ is positive (as defined in the caption of Fig. \ref{fig2}), we can determine the intervals of the temperature $T$ and the chemical potential $\mu$. The results are presented in Table \ref{tab3}.

\subsection{\label{sec2-3}Application of MC to the Parameter Set ($T$, $\varepsilon$, $\mu$, and $y$) (Case $\varepsilon \ne 0$)}
\begin{figure*}[h]
\begin{center}
\includegraphics[height=58mm]{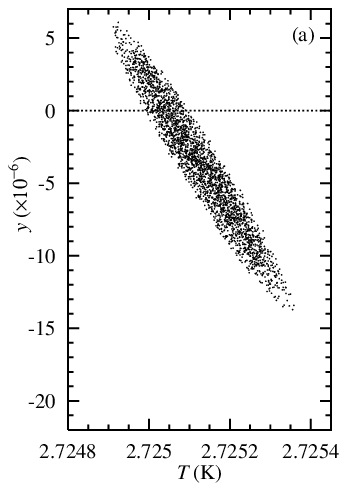}
\includegraphics[height=58mm]{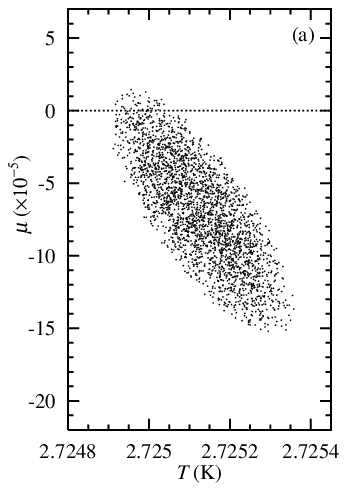}
\includegraphics[height=58mm]{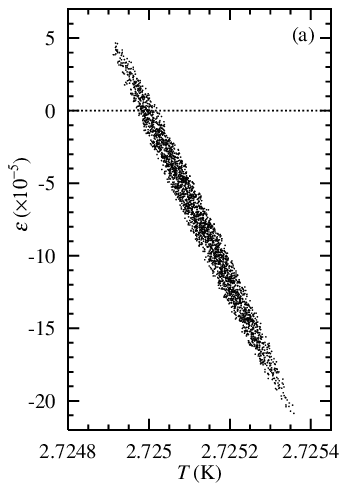}\\
\includegraphics[height=50mm]{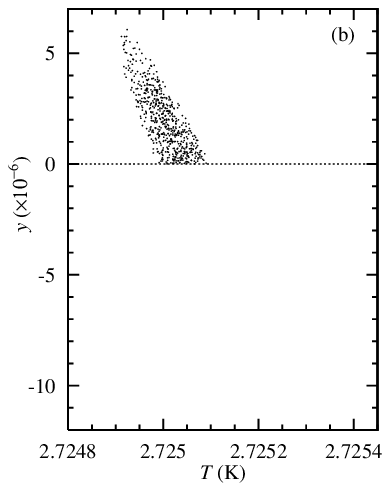}
\includegraphics[height=50mm]{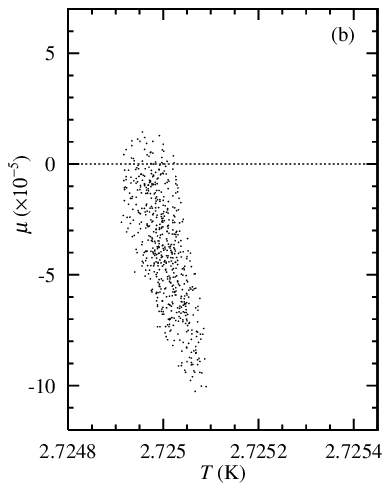}
\includegraphics[height=50mm]{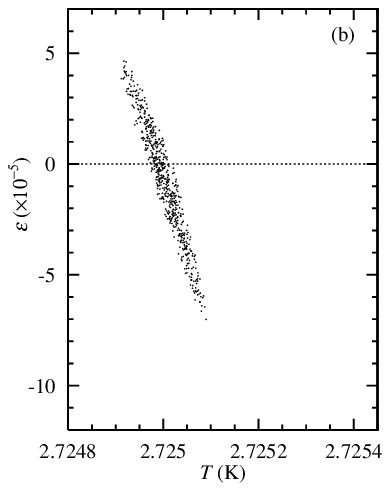}
\end{center}
\caption{\label{fig3}Parameter ensemble ($T,\; \varepsilon,\; \mu,\ y$) with $\chi^2<\chi_{\rm min}^2 (44.65) + \Delta \chi^2 (0.67)$. Number of events generated is $3\times 10^6$ (3M). (a) $N = 2699$. (b) $N_{y>0} = 597$.}
\end{figure*}

Inserting the space distortion $d=3+\varepsilon$ into the above calculations yields the distributions shown in Fig. \ref{fig3}. Numbers of events satisfying $y > 0$ for the two calculations (assuming isotropic and distorted space) are listed in Table \ref{tab3}.
\begin{table*}
\begin{center}
\caption{\label{tab3}Averages of parameter ensembles ($N$($y$ all) and $N_{y>0}$ events) assuming $\chi^2 < \chi_{\rm min}^2 + \Delta \chi^2$ in Figs. \ref{fig2} and \ref{fig3}}
\begin{tabular}{@{}lccccc@{}}
\tableline\tableline
Calc. & $\!\!\!\!\!\!\!\!\! y({\rm all\: and} >0)\;(\times 10^{-6})\!\!\!\!\!\!\!\!\!\!\!$  & $T$ (K) & $\mu\;(\times 10^{-5})$ & $\varepsilon\;(\times 10^{-5}$) & $\langle \chi^2\rangle$\\
\tableline
I ($N=$1529) & $1.682\pm 2.138$ & $2.72499\pm 0.00002$ & $-2.707\pm 2.477$ & --- & $45.47$\\
I ($N_{y>0}=$1142) & $2.627\pm 1.551$ & $2.72499\pm 0.00002$ & $-3.5879\pm 2.098$ & --- & $45.45$\\
\tableline
II ($N=$2699) & $-3.674\pm 4.178$ & $2.72513\pm 0.00009$ & $-6.883\pm 3.563$ & $-7.578\pm 5.462$ & $45.09$\\
II ($N_{y>0}=$597) & $2.001\pm 1.367$ & $2.72500\pm 0.00004$ & $-3.885\pm 2.585$ & $-0.775\pm 2.497$ & $45.14$\\
\tableline
\end{tabular}
\end{center}
\end{table*}

\section{\label{sec3}Analyses of Residual Spectrum in terms of Allowed Parameter Combination Ensembles}
\begin{figure}[h]
\begin{center}
\includegraphics[height=58mm]{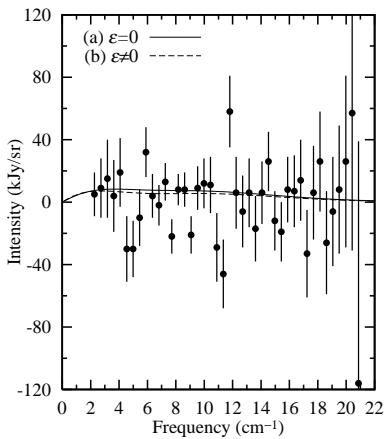}
\end{center}
\caption{\label{fig4}Analysis of COBE residual spectrum for allowed combinations of parameter ensembles $(\varepsilon,\ T,\ \mu,\ y>0)$. (a) $T=2.72499$ K, $\mu = -3.5879\times 10^{-5}$ and $y=2.627\times 10^{-6}$ (see Table \ref{tab3}), and $\chi^2 = 50.2$. (b) $T=2.72500$ K, $\varepsilon=-0.775\times 10^{-5}$, $\mu = -3.885\times 10^{-5}$ and $y=2.001\times 10^{-6}$ (see Table \ref{tab3}), and $\chi^2 = 47.8$.}
\end{figure}

If $|\mu|\ll 1$, Eq. (\ref{eq7}) is simplified, and the residual spectrum becomes
\begin{eqnarray}
&\!\!\!\!\!&\!\!\!\!\! U^{(\rm residual\ spectrum)} \nonumber\\
&\!\!\!\!\! = &\!\!\!\!\! N_c(d)\cdot\left(\frac{\nu}c\right)^d\nonumber\\
&\!\!\!\!\!&\!\!\!\!\! \times\left[\frac{1}{e^{(x+\mu)}-1} - \frac{1}{e^{x}-1} + \frac{yxe^x}{(e^x-1)^2}\left(x\coth \frac x2-4\right)\right]\nonumber\\
&\!\!\!\!\! \cong &\!\!\!\!\! N_c(d)\cdot\left(\frac{\nu}c\right)^d\nonumber\\
&\!\!\!\!\!&\!\!\!\!\! \times\left[-\mu\frac{e^x}{(e^x-1)^2} + \frac{yxe^x}{(e^x-1)^2}\left(x\coth \frac x2-4\right)\right].
\label{eq9}
\end{eqnarray}
Computing the residual spectrum by means of Eq. (\ref{eq9}) yields Fig. \ref{fig4}, in which the allowed combinations of parameters ensembles are evaluated for $\varepsilon = 0$ and $\varepsilon \ne 0$. This figure reveals slight difference between the residual spectra for $\varepsilon = 0$ and $\varepsilon = -0.755\times 10^{-5}$.

\section{\label{sec4}Concluding Remarks and Discussions}
\begin{figure}[h]
\begin{center}
\includegraphics[height=52mm]{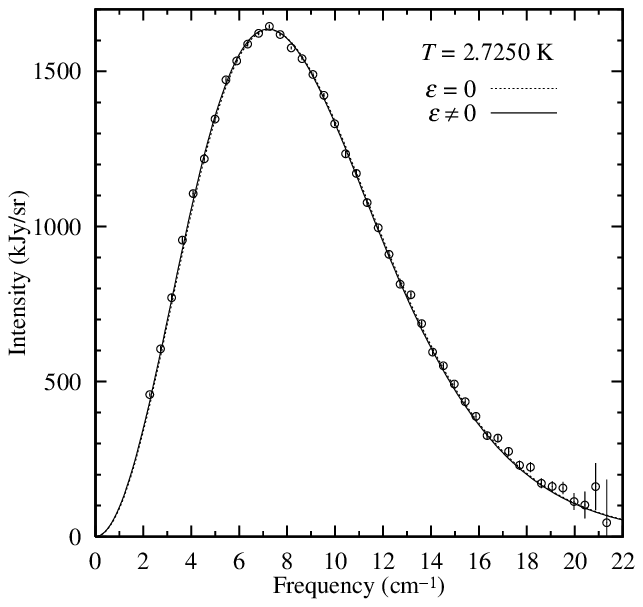}
\end{center}
\caption{\label{fig5}Analysis of COBE dipole spectrum at fixed $T$ ($2.7250$ K), computed from  Eq. (\ref{eq16}).}
\end{figure}

\noindent
{\bf (C1) } We have computed the distortion of the space-dimension $\varepsilon$ in the NASA COBE monopole data \citep{Fixsen2002,COBE2005} using Eqs. (\ref{eq-3}) and (\ref{eq2}). Difference between the two results is attributed to numerical difference in the $hc/k_B$ values. The magnitude of our estimation of $\varepsilon$ is $(3\pm 25)\times 10^{-6}$, should be compared with that of \citet{Caruso2009} (namely, $-(9.6\pm 0.1)\times 10^{-6} \approx -1\times 10^{-5}$). As seen in third row of Table \ref{tab1}, when we adopt the same numerical value for $hc/k_B$ as \citet{Biyajima2012} in Eq. (\ref{eq-3}), we obtained $\varepsilon = (-53\pm 106) \times 10^{-6}$ $(\sigma)$ (and $|\varepsilon| \leq 1.74\times 10^{-4}$ $(2\sigma)$). It seems to be inadequate to tune the physical constant factor $hc/k_B$ as \citet{Caruso2009}.
\medskip\\
{\bf (C2) } To extend our investigation of the distortion of the space-dimension in the COBE data, we propose that chemical potential $\mu$ and the SZ effect $y$, in addition to $\varepsilon$, be included in Eq. (\ref{eq7}). Applying Eq. (\ref{eq7}) to the COBE monopole data and the MC method to parameters sets ($T$, $\varepsilon$, $\mu$, and $y$) and ($T$, $\varepsilon = 0$, $\mu$, and $y$), we have obtained the allowed parameter ensembles, assuming the positivity of the SZ effect $y$. The following limits of the parameters have been obtained at $T = 2.72500\pm 0.00004$ K;
\begin{eqnarray}
&\!\!\!\!\!\!&\!\!\!\!\!\! |\varepsilon| < 6\times 10^{-5}\ (2\sigma)\ (\varepsilon = (-0.78\pm 2.50)\times 10^{-5}\ (\sigma)),
\nonumber\\
&\!\!\!\!\!\!&\!\!\!\!\!\! |\mu| < 9\times 10^{-5}\ (2\sigma)\ (\mu = (-3.9\pm 2.6)\times 10^{-5}\ (\sigma)), \nonumber\\
&\!\!\!\!\!\!&\!\!\!\!\!\! |y| < 5\times 10^{-6}\ (2\sigma)\ (y = (2.0\pm 1.4)\times 10^{-6}\ (\sigma)).
\label{eq10}
\end{eqnarray}
Assuming no distortion of the space-dimension, i.e., $\varepsilon= 0$, the following limits are obtained at $T = 2.72499\pm 0.00002$ K,
\begin{eqnarray}
&\!\!\!\!\!\!&\!\!\!\!\!\! |\mu| < 8\times 10^{-5}\ (2\sigma),\vspace{-3mm}\nonumber\\
&\!\!\!\!\!\!&\!\!\!\!\!\!  |y| < 6\times 10^{-6}\ (2\sigma).
\label{eq11}
\end{eqnarray}
The above limit of $|\mu|$ is almost identical to that reported by \citet{Fixsen2002} ($|\mu|<9\times 10^{-5}$). However, our estimated limit of $|y|$ is smaller than that of these authors ($|y|<1.2\times 10^{-5}$), probably because SZ effect $|y|$ was assumed as positive in our study. Moreover, it is interesting that limits $|\mu|$ and $|y|$ slightly depend on the magnitude of distortion $|\varepsilon|$. Allowing for the distortion of the space-dimension, among three limits we observe the following order:
\begin{eqnarray}
|\mu|\gesim |\varepsilon| > |y|.
\label{eq12}
\end{eqnarray}
{\bf (C3) } According to \citet{Durrer2008}, as the Universe is re-ionized at some redshift $z_{ri}$, during the ionization process, the electrons gain kinetic energy (estimated at around $10$ eV). Denoting the electron temperature at the re-ionization by $T_{ri}$, we obtain the following expression for the SZ effect:
\begin{eqnarray}
y &\!\!\!=&\!\!\! \frac{\sigma_TT_{ri}n_e(t_0)}{m_eH_0\sqrt{\Omega_m}(Z_{ri} + 1)^2}\int_0^{z_{ri}}dz(z+1)^{5/2} \nonumber\\
 &\!\!\!=&\!\!\! 5\times 10^{-7} \frac{\Omega_b h^2}{\sqrt{\Omega_m h^2}}(z_{ri} + 1)^{3/2} \left(\frac{T_{ri}}{10\ {\rm eV}}\right),
\label{eq13}
\end{eqnarray}
where $\Omega_m$, $\Omega_b$, and $h$ are the matter density, the baryon density, and the scale parameter, respectively. Given $\Omega_m h^2 \cong 0.13$, $\Omega_b h^2 \cong 0.022$, the re-ionization redshift for the limit $y < 10^{-5}$ is $z_{ri} < 50(T_{ri}/10\ {\rm eV})^{-2/3}$. The limit $y < 10^{-6}$ implies that
\begin{eqnarray}
z_{ri} < 10\left(\frac{T_{ri}}{10\ {\rm eV}}\right)^{-2/3}.
\label{eq14}
\end{eqnarray}
This set of parameters ($T_{ri}\cong 10$ eV, and $z_{ri}\sim 10$) yields $y\sim 10^{-6}$, which is consistent with the limits in Eqs. (\ref{eq10}) and (\ref{eq11}).
\medskip\\
{\bf (D1) } For the sake of reference, we analyze the same data by means of the relativistic formula denoted by $U_{\rm SZ(*)}$ \citep{Itoh1998,Challinor1998}
\begin{eqnarray}
 &\!\!\!\!\!\!&\!\!\!\!\!\! U_{\rm SZ(*)} (T,\: x,\: y_*,\: \theta_e) \nonumber\\
 &\!\!\!\!\!\!&\!\!\!\!\!\! = C_B \frac{y_* \, \theta_{e} x e^{x}}{(e^{x}-1)^2} \, \left[  \,
Y_{0} \, + \, \theta_{e} Y_{1} \, + \, \theta_{e}^{2} Y_{2} \, + \,  \theta_{e}^{3} Y_{3} \, + \,  \theta_{e}^{4} Y_{4} \, \right], \nonumber\\
\label{eq15}
\end{eqnarray}
where $y_* = \int dl n_e \sigma_T$, $\theta_e = k_BT_e/m_ec^2$, $Y_0 = x\coth(x/2) - 4$, and
\begin{eqnarray*}
Y_1 &\!\!\!=&\!\!\! - 10 + \frac{47}2x\coth\frac x2 - \frac{42}5x^2\coth^2\frac x2 + \frac{7}{10}x^3\coth^3\frac x2 \nonumber\\
&\!\!\! &\!\!\! + \frac{x^2}{\sinh^2(x/2)}\left( - \frac{21}5 + \frac 75 x\coth\frac x2\right).
\end{eqnarray*}
$Y_2$, $Y_3$ and $Y_4$ are explicitly given in \citet{Itoh1998}. According to descriptions on temperature of electron $T_e$'s in \citet{Weinberg2008,Naselsky2006}, the temperatures of the ionized plasma in cluster of galaxies is approximately on $(6.5\sim 7)\times 10^6$ K. Our results are presented in Table \ref{tab4}. It is shown that estimated values are almost the same in Table \ref{tab2}. (Notice that ``$y_*\theta_e$'' in Table \ref{tab4} corresponds to ``$y$'' in Table \ref{tab2}.) 
\begin{table*}
\begin{center}
\caption{\label{tab4}Analysis of COBE monopole spectrum by means of Eq. (\ref{eq15}) and $\theta_e =$ 0.6, 0.8 and 1.0 keV (Fixed)} 
\begin{tabular}{@{}ccccccc@{}}
\tableline\tableline
 $T_e$ (keV) & $T$ (K) & $\mu\;(\times 10^{-5})$ & $y_*\;(\times 10^{-3})$ & $y_*\theta_e\;(\times 10^{-6})$ & $\varepsilon\;(\times 10^{-5})$ & $\chi^2$\\
\tableline
 $0.6$ & $2.72500\pm 0.00004$ & $-2.6\pm 5.6$ & $1.4\pm 4.1$
 & $1.6\pm 4.8$ & --- & $44.9$ \\
 $0.8$ & $2.72500\pm 0.00004$ & $-2.6\pm 5.6$ & $1.1\pm 3.1$
 & $1.7\pm 4.9$ & --- & $44.9$ \\
 $1.0$ & $2.72500\pm 0.00004$ & $-2.6\pm 5.6$ & $0.9\pm 2.5$
 & $1.7\pm 4.9$ & --- & $44.9$ \\
 $0.6$ & $2.72512\pm 0.00028$ & $-6.9\pm 10.7$ & $-3.1\pm 10.7$
 & $-3.7\pm 12.5$ & $-7.5\pm 16.3$ & $44.7$ \\
 $0.8$ & $2.72512\pm 0.00029$ & $-6.9\pm 11.1$ & $-2.4\pm 8.3$
 & $-3.7\pm 13.0$ & $-7.5\pm 17.0$ & $44.7$ \\
 $1.0$ & $2.72513\pm 0.00028$ & $-6.9\pm 10.7$ & $-1.9\pm 6.5$
 & $-3.8\pm 12.6$ & $-7.6\pm 16.3$ & $44.7$ \\
\tableline
\end{tabular}
\end{center}
\end{table*}
\medskip\\
{\bf (D2) } The dipole spectrum can be analyzed by the derivative of the Planck distribution \citep{Fixen1996,Henry1968,Sugiyama2001}, given as:
\begin{eqnarray}
 &\!\!\!&\!\!\! \diff{U_{\rm Planck}}{T} \delta T = C_B\cdot\left(\frac{\nu}c\right)^3 \frac{x e^x}{(e^x-1)^2} \frac{\delta T}T,\nonumber\\
&&\ \ \ \ \mapright{\ \delta T\to T_{\rm amp}\ }\ N_c(d)\cdot\left(\frac{\nu}c\right)^d \frac{x e^x}{(e^x-1)^2} \frac{T_{\rm amp}}T.
\label{eq16}
\end{eqnarray}
The results of this analysis are provided in Table \ref{tab5} and Fig. \ref{fig5}. As seen in Table \ref{tab5}, the $\chi^2$ is improved by imposing a finite $\varepsilon = d-3$ on the system. From the dipole amplitude $T_{\rm amp} = 3.597\pm 0.044$ mK, the velocity of our solar system (with the Universe fixed) is estimated as $v = 396$ km s$^{-1}$, about $6.6$ \% higher than the estimate of \citet{Fixen1996}. The anomalously large value of $\varepsilon$ obtained in this analysis requires further investigation.
\begin{table}
\begin{center}
\caption{\label{tab5}Analysis of COBE dipole spectrum by Eq. (\ref{eq16}) at $T = 2.7250$ K} 
\begin{tabular}{@{}cccc@{}}
\tableline\tableline
$T$ (K) & $T_{\rm amp}$ (mK) & $\varepsilon\;(\times 10^{-2})$ & $\chi^2$ \\
\tableline
$2.7250$ (fixed) & $3.380\pm 0.004$ & --- & $118.2$ \\
$2.7250$ (fixed) & $3.597\pm 0.044$ & $-2.0\pm 0.4$ & $92.5$ \\
\tableline
\end{tabular}
\end{center}
\end{table}
\medskip\\
{\bf (D3) } From Eq. (\ref{eq2}), a modified Stefan-Boltzmann law is derived as,
\begin{eqnarray}
 \int_0^{\infty} \frac{x^d}{e^x-1}dx \approx \Gamma (d+1) \left[\zeta (4) - \varepsilon \sum_{n=1}^{\infty}\frac{\ln n}{n^4}\right],
\label{eq-18}
\end{eqnarray}
where $\zeta (4)$ is the Riemann's zeta-function and the second term
  $\sum_{n=1}^{\infty}\ln n/{n^4} = 9.89113\times 10^{-2}$. Through Eq. (\ref{eq-18}), we are able to discuss the problem of the distortion of the space-dimension in a future in more detail, provided that the error bar of $\varepsilon$ becomes smaller.

\acknowledgments

The authors sincerely thank Prof. J. C. Mather and Prof. D. J. Fixsen for their kindness in supplying NASA COBE data and discussing analysis methods during the early stages of this investigation. Subsequently, one of the authors (M. B.) would like to thank Prof. N. Sugiyama for insightful conversations. Moreover, it is our pleasure to acknowledge Dr. F. Caruso and Dr. V. Oguri for our communications with them.

\end{document}